\documentstyle[12pt]{article}
\newcommand{\be}{\begin{equation}}
\newcommand{\ee}{\end{equation}}
\newcommand{\al}{\mbox{$\alpha$}}

\newcommand{\s}{\mbox{$\sigma$}}

\newcommand{\bi}[1]{\bibitem{#1}}
\newcommand{\fr}[2]{\frac{#1}{#2}}

\newcommand{\gm}{\mbox{$\gamma_{\mu}$}}

\newcommand{\gf}{\mbox{$\gamma_{5}$}}

\newcommand{\Ima}{\mbox{Im}}

\begin{document}

\normalsize
\begin{flushright}{BINP 95-59\\ UQAM-PHE/95-09\\ July 1995}
\end{flushright}
\vspace{0.5cm}
\begin{center}{\Large \bf Pure electroweak mechanism for the
electric dipole moment of neutron in the Kobayashi-Maskawa
model.}\\
\vspace{1.0cm}
{\bf C. Hamzaoui}\footnote{E-mail: hamzaoui@mercure.phy.uqam.ca}\\
D\'epartement de Physique, Universit\'e du Qu\'ebec \`a Montr\'eal,\\
Case Postale 8888, Succ. Centre-Ville, Montr\'eal, Qu\'ebec, Canada, H3C 3P8.

\vspace{0.7cm}

and\\
\vspace{0.7cm}

{\bf  M.E. Pospelov}\footnote{E-mail: pospelov@inp.nsk.su}\\
Budker Institute of Nuclear Physics, 630090 Novosibirsk, Russia
\vspace{3.5cm}
\end{center}

\begin{abstract}
The pure electroweak three-loop mechanism for the induced
electric and chromoelectric dipole moments of quarks is studied
in the Kobayashi-Maskawa model with three and four generations.
In the standard three generation case, this mechanism is found
to produce a negligible contribution to the electric dipole moment
of neutron. In the presence of the fourth heavy generation, however,
pure electroweak corrections are important and might be several times
larger than the corresponding QCD contribution for the masses of
heaviest quarks $\sim$ 500-600 GeV. The resulting
electric dipole moment of neutron naturally arises at the level
of $10^{-29} \,e\cdot cm$. The effects of the fourth generation
physics are parametrized at standard electroweak scale by the
presence of the effective charged right-handed currents.

\end{abstract}
\newpage
\section{Introduction}
In this letter, we consider the pure electroweak three loop contribution
to the neutron electric
dipole moment in the Kobayashi-Maskawa (KM)
model with four generations. Besides its
interesting predictions for B-meson physics \cite{HSS}, the
enlarged variant of the KM model leads to the enhancement of the neutron
electric dipole moment (EDM) in comparison with Standard Model
case \cite{HamPos}.
This enhancement is linked with the short distance contribution to
the electric and chromoelectric dipole moments (CEDM) of quarks.

The origin of CP-violation in the Kobayashi-Maskawa (KM) model resides in
the complexity of the quark mixing matrix. This requires four
semi-weak vertices to generate a flavour-diagonal CP-odd
amplitude. It was shown that the EDMs and CEDMs of quarks cannot be generated
at the lowest possible two-loop order
\cite{Shab}. Thus, the QCD corrections are brought in to
prevent these quantities from the identical cancellation.
The detailed study of these operators at three loop order (two
electroweak plus one gluonic) was done in
SM by Khriplovich \cite{Kh} and by us for its four generation
modification \cite{HamPos}.

Here, we replace the hard gluon loop by another one of electroweak
in the EDM inducing graphs.
Specifically, we concentrate on the
large renormalization factor of the axial coupling of Z-boson
with fermion proportional to $m_t^2$ in SM
and its relevance for the induced EDM in the presence
of an extra heavy generation of quarks \cite{NOV}.

The purpose of this work is to compute EDMs and CEDMs at three
loop electroweak order and then compare our results with
corresponding QCD induced values \cite{HamPos,Kh}.

\section{Electroweak corrections to EDM}
At first glance the problem of this calculation appears
to be very complicated.
However, taking into account the explicit mass hierarchy in this problem,
we reduce the three-loop calculation
to one-loop integrals.
We assume that:
\be
m_h^2;\;m_g^2\gg m_t^2\gg m_w^2\gg m_i^2,
\label{eq:scale}
\ee
where we have denoted the heaviest quarks as h and g; i
represents the standard set of "light" flavours: u, d, s, c and b.
The first inequality is assumed in order to single
out parametrically the contributions proportional to $m_h^2$ or
$m_g^2$.

The typical representatives of the diagrams
to be calculated are depicted in Fig. 1. The solid line represents
the fermions; wavy lines are charged electroweak bosons
and the dashed line are the neutral ones. The position of an
external photon or gluon is not indicated here.

In the following calculation, we consider the EDM and CEDM
operators of the strange quark. Then the arrangement of flavours
along the fermion
line is determined in SM uniquely by \cite{Kh}:
\be
i\Ima(V^*_{ts}V_{td}V^*_{cd}V_{cs})
s[t(b-d)u-u(b-d)t+u(b-d)c-c(b-d)u+c(b-d)t-t(b-d)c]s.
\label{eq:FS1}
\ee
The enlarged KM matrix possesses in general three independent
CP-odd invariants, one of which being
distinguished by the dynamical
enhancement \cite{HamPos}:
\be
i\Ima(V^*_{ts}V_{tb}V^*_{hb}V_{hs})
s[t(b-g)h-h(b-g)t+U(b-g)t-t(b-g)U+h(b-g)U-U(b-g)h]s.
\label{eq:FSfin}
\ee
The capital U here denotes the propagation of u and c quarks which we
are free to regard massless and degenerate inside the loops.
This degeneracy is the factor which leads to the identical cancellation of
the amplitude (\ref{eq:FS1}) in the SM. Therefore (\ref{eq:FS1})
is forced to be
proportional to $m_c^2$ whereas (\ref{eq:FSfin}) is determined
by heavy mass scale like $m_t^2$ \cite{HamPos}. This is the main
source of the EDM enhancement in the KM model with four
generations. The additional source of the enhancement probably lies in
the numerical significance of the CP-odd combination
$\Ima(V^*_{ts}V_{tb}V^*_{hb}V_{hs})$ which could naturally reach
the level of $\lambda^5$, where $\lambda$ is the Wolfenstein
parameter \cite{HSS,HamPos}.

We begin from the shortest distances and calculate first the
effective one-loop induced flavour changing neutral currents
"electroweak penguin", which is well known from the
kaon physics:
\be
{\cal L}^{(1)}=\sum_{i,j,f}a_jV_{ij}V^*_{fj}\bar{q}_i\gm(1-\gf)
 q_fZ_\mu
\ee
To a good accuracy, it is sufficient to take only leading
contribution to the coefficients $a_j$ which are proportional
to the square of the heaviest masses:
\be
{\cal L}^{(1)}=\fr{g_w}{\cos\theta_W}\fr{\al_w}{16\pi}Z_\mu\left[
\fr{m_h^2}{m_w^2}\sum_{i,f}V_{hi}V^*_{hf}\bar{q}_f\gm\fr{(1-\gf)}{2}
 q_i-
\fr{m_g^2}{m_w^2}\sum_{i',f'}V_{i'g}V^*_{f'g}\bar{q}_{f'}\gm\fr{(1-\gf)}{2}
q_{i'}\right],
\label{eq:eff1}
\ee
in analogy to the SM expression with dependence on $m_t^2$.
The heavy mass dependence here originates from the longitudinal
part of W-boson propagator or from the equivalent graph with charged
non-physical higgs boson \cite{NOV}.
Taking the new effective vertex
generated by (\ref{eq:eff1}), we reduce the remaining computation to
the one
presented in Fig. 2.

The second step is to integrate out the neutral
boson, which could be done along the same line. At
this point, however, it is useful to treat the three and four
generation models separately.

In the four generation case, the characteristic
loop momenta are large, which allow us again to take only the longitudinal
part of the Z-boson propagator. This leads us
to obtain effective charged right handed
currents in an easy way. Taking into account the flavour structure
in (\ref{eq:FSfin}), we obtain the effective lagrangian for the s -
t transition:
\be
{\cal L}^{(2)}\simeq -\fr{g_w}{\sqrt{2}}V_{tb}V^*_{hb}V_{hs}
(\fr{\al_w}{16\pi})^2
\fr{m_tm_s}{m_w^2}\fr{(m_h^2+m_g^2)}{m_w^2}\log(\fr{m_{h(g)}^2}{m_t^2})
\bar{q}_t\gm\fr{(1+\gf)}{2}q_sW^+_\mu +(h.c.),
\label{eq:eff2}
\ee
where we have omitted all constants in comparison with "large"
logarithmic factors. This logarithmic accuracy is motivated by
theoretical considerations. However, in the final numerical result,
we set all logarithms to
unity. This allows us to avoid the problem of a true
two-loop calculation which is not reducible to two independent
integrations. It is important to remember that the vertices
(\ref{eq:eff1}) and (\ref{eq:eff2}) are indeed effective and do
not survive if the incoming momenta are larger than all masses
of particles flowing inside the loops. For this reason, the upper limit
for logarithmic integral over the loop momentum coincides with
$m_h$ if $m_h<m_g$ and with $m_g$ if $m_g<m_h$. The analysis of
the precision electroweak data suggests that h and g quarks
must be sufficiently degenerate in masses. From here to below,
we put $m_g \simeq m_h$. It is worth to note also the constructive
interference between the two terms in (\ref{eq:eff2}) proportional
to $m_g^2$ and $m_h^2$, in contrary to the mass dependence of
the electroweak parameter $\rho$ \cite{NOV}. The contribution of
the fourth generation to $\rho$ vanishes at $m_h=m_g$.

The situation is quite different in the SM. Due to the presence of
right-handed currents in the s-c or d-c transition, those transition
amplitudes are suppressed not
only by $m_{s(d)}$ and $m_c$ but also by the factor
$m_b^2/m_Z^2$ reflecting the GIM property of
(\ref{eq:FS1}). Taking into account that the result for EDM of
s or d quark must be
proportional to $m_c^2$, although the interchange of a gluon loop by
an electroweak one replaces
$\al_s(q^2\simeq m_b^2)$ by
$\al_w\fr{m_t^2m_b^2}{m_w^4}$, we deduce that the total effect is
negligibly small.

The presence of right-handed currents itself does not
necessarily imply CP-violation. The latter arises through the
complex phase of $V_LV^{\dagger}_R$, the product of right- and left-handed
KM matrices,
\be
{\cal L}=\fr{g_w}{2\sqrt{2}}W^+_\mu\sum_{f,i}
\left[\bar{q}_f(V_L)_{fi}\gm (1-\gf)q_i+
\bar{q}_f(V_R)_{fi}\gm (1+\gf)q_i\right]+(h.c.),
\ee
and is known to exist even in two generations. This
is exactly what happens at the last stage of our calculation.
EDM and CEDM of s-quark results from the mixing between second
and third generations in the presence of right-handed currents
given by (\ref{eq:eff2}) depicted in Fig. 3. The results are:
\begin{eqnarray}
\label{eq:EDM}
d_s=-\fr{5e}{3}\Ima(V^*_{ts}V_{tb}V^*_{hb}V_{hs})
(\frac{\alpha_w}{4\pi})^2\frac{1}{16\pi^2}\fr{G_F}{\sqrt{2}}m_s
\frac{m_t^2m_h^2}{4m_w^4}\log(\fr{m_{h}^2}{m_t^2})\\
\tilde{d}_s=-g_s\Ima(V^*_{ts}V_{tb}V^*_{hb}V_{hs})
(\frac{\alpha_w}{4\pi})^2\frac{1}{16\pi^2}\fr{G_F}{\sqrt{2}}m_s
\frac{m_t^2m_h^2}{4m_w^4}\log(\fr{m_{h}^2}{m_t^2}),
\label{eq:CEDM}
\end{eqnarray}
where $d_s$ and $\tilde{d_s}$ are determined as the coefficients
in front of $\fr{1}{2}\bar{q}(F\s)\gf q$ and
$\fr{1}{2}\bar{q}t^a(G^a\s)\gf q$ respectively. Due to
the inequality (\ref{eq:scale}), we have taken into account only the
longitudinal part of W-propagator. The use of longitudinal
parts of gauge boson propagators throughout the calculation of EDM in
four generation case maximizes the size of CP-violation. This
also means that the obtained result arises entirely from the Higgs sector
phenomenon. Then, the whole calculation could be performed in
the t'Hooft-Feynman gauge and therefore only scalar bosons should be taken
into account. Thus, we can rewrite the results given in
(\ref{eq:CEDM}) and (\ref{eq:EDM}) in terms of Yukawa couplings
$f_i=m_i/v$ of heavy quarks:
\be
\fr{\tilde{d}_s}{g_s}=\fr{3d_s}{5e}=-\Ima(V^*_{ts}V_{tb}V^*_{hb}V_{hs})
\frac{1}{1024\pi^6}\fr{G_F}{\sqrt{2}}m_s
f_t^2f_h^2
\log(\fr{f_{h}^2}{f_t^2}),
\label{eq:ans}
\ee
where $v=246 $GeV is the vacuum expectation value of the scalar
field.

The extraction of the EDM of neutron resulting from the effective
interaction (\ref{eq:CEDM}) depends on our understanding of
low-energy hadronic physics. Based on the chiral perturbation
estimation proposed in \cite{KKZ}, we find the EDM of neutron
at the level:
\be
\label{eq:NEDM}
d_N\sim
e\Ima(V^*_{ts}V_{tb}V^*_{hb}V_{hs})f_t^2f_h^2\cdot
2\cdot10^{-26}\;cm
\ee
The comparison of (\ref{eq:NEDM}) with the corresponding
QCD-induced contribution to EDM obtained earlier in
Ref.\cite{HamPos} shows, in principle, the same order of magnitude
for both results. In the most optimistic scenario about the
values of CP-odd phase invariant, combined with the masses of heavy
quarks around $500$GeV, we obtain the neutron EDM to arise at the
level:
\be
d_N\sim10^{-29}\;e\cdot cm.
\label{eq:num}
\ee
The main source of the numerical smallness is connected with
the tiny numerical coefficient in expression (\ref{eq:ans}).

\section{ Discussions and Conclusions}
We would like to point out
that there exists another contribution to the EDM of neutron
from the physical Higgs
boson loop which should be taken into account as well.  One may
undertake that calculation along the same line and obtain the
effective right-handed currents as well. This means that the
value of EDM depends not only on unknown $m_g$ and $m_h$ but also on the
mass of real Higgs boson. To our logarithmic
accuracy, these corrections are unimportant if we believe that
$m_{Higgs}$ is also very large, somewhere around $m_{h(g)}$.

The last diagram which could contribute to EDM at this order is
the "rainbow" graph, Fig. 4, where all wavy lines are W-bosons.
It could be checked, however, that even if this diagram provides
any nonvanishing contribution, it cannot generate
an $m_{h(g)}^2$-dependence in the result.

Our estimate shows that the electroweak contributions to the EDM of
neutron are comparable with QCD ones in the case of four
generations and negligibly small in SM. The QCD effects dominate
over pure electroweak contribution at $m_{h(g)}\sim m_t$, whereas
at $m_{h(g)}\sim 500-600$GeV the latter dominates. The numerical result
(\ref{eq:num}) then could be regarded as the maximal value of
EDM which can be obtained through the Kobayashi-Maskawa type of
CP-violation in the presence of an additional heavy generation.
Despite the significant enhancement in comparison with SM case,
it is too low to be detected even at the future generation
of experiments aiming at the search of EDM. This implies also
that the reliable information and limits on the parameter of the
model with four generations may come only from the analysis of
the electroweak precision data and K, B-meson physics
\cite{HSS}. The latter case
requires a further analysis on the role of large electroweak
radiative corrections.

As a concluding remark, we have emphasized the main mechanism
for the induced EDM of neutron in the hypothetical case of two or
more additional heavy generations, (h, g); (h', g');..., with
large mixing between them. Then the maximum of CP-violation at
low energies comes from Weinberg operator \cite{Wein} which is
known to exist in KM model at the lowest possible three-loop
order \cite{Posp}. In the four generation case, this operator is
suppressed by the factor $m_b^2/m_w^2$ \cite{HamPos}, which
disappears at five or more generations. At the same time, there
is no severe limits on the mixing between heavy generations from
the low energy phenomenological data and one may expect the
corresponding CP-odd invariant, $\Ima(V^*_{tg'}V_{tb}V^*_{hb}V_{hg'})$
to be large.
\eject \hfill

\begin{flushleft}{\Large\bf Acknowledgments}\end{flushleft}
We thank I.B. Khriplovich, V.N. Novikov and M.I.
Vysotsky for attracting our attention to the significance of
electroweak radiative corrections to EDM and for helpful
discussions. We thank also F. Boudjema for his kind advise
and comments. The work of C.H. was partially funded by N.S.E.R.C.
of Canada, and the research of M.P. was supported in part by the
I.C.F.P.M., INTAS grant \#\,93-2492.

\end{document}